\documentclass[twocolumn,showpacs,preprintnumbers,amsmath,amssymb]{revtex4}
\usepackage{graphicx}
\usepackage{dcolumn}
\usepackage{bm}

\begin{document}

\title{Local temperature and chemical potential inside a mesoscopic device driven out of equilibrium}

\author{Pei Wang}
\email{wangpei@zjut.edu.cn}
\affiliation{Institute of applied physics, Zhejiang University of Technology, Hangzhou, P. R. China
}

\date{\today}

\begin{abstract}
In this paper we introduce a method of calculating the local temperature and chemical potential inside a mesoscopic device out of equilibrium. We show how to check the conditions of local thermal equilibrium as the whole system is out of equilibrium. Especially we study the onsite chemical potentials inside a chain coupled to two reservoirs at a finite voltage bias. In the presence of disorder we observe a large fluctuation in onsite chemical potentials, which can be suppressed by the electron-electron interaction. By taking average with respect to the configurations of disorder, we recover the classical picture where the voltage drops monotonously through the resistance wire. We prove the existence of local intensive variables in a mesoscopic device which is in equilibrium or not far from equilibrium.
\end{abstract}


\maketitle

\section{introduction}

In recent years the thermodynamic properties of a quantum system driven out of equilibrium have attracted more and more attention~\cite{rigol07,moeckel,kollath,barthel08,gemmer,znidaric,rigol08}. Especially, in mesoscopic transport a standard method of studying nonequilibrium steady states (NESS) has been established by using the evolution approach in Keldysh formalism~\cite{datta,jauho,hershfield93}. In mesoscopic transport we study small quantum systems with a few degrees of freedom which must be described by quantum mechanics. The small system is connected to several infinite reservoirs which are in thermal equilibrium with different temperatures and chemical potentials. The nonequilibrium steady state is approached by an evolution starting from the supposed initial condition. The evolution approach has been widely used to study the transport through quantum point contacts, quantum dots, single molecules and carbon nanotubes. Good agreements between theoretical predictions and experimental results have been obtained. 

Now a fundamental question arises as to whether the concepts in equilibrium thermodynamics apply to the mesoscopic system driven out of equilibrium. For example, temperature is an important concept in equilibrium thermodynamics, where it is also called the intensive thermodynamic variable of a system. According to the zeroth law of thermodynamics, it is always possible to assign a temperature to a system in thermal equilibrium. Even if the whole system is out of equilibrium, when the conditions of local equilibrium are fulfilled, we can divide the system into small cells which are in thermal equilibrium to a good approximation. In principle we are then able to define the local temperature in each cell. However, up to now the sole way of calculating temperature according to the density matrix of a system is by using the extremal principle in statistical mechanics. According to Glansdorff and Prigogine~\cite{glansdorff} in systems away from equilibrium there is no general variational principle. Then temperature, arising from the Lagrangian multiplier in statistical mechanics, loses its meaning. Even if many authors make their efforts in the local extremal principles~\cite{glansdorff,onsager31,prigogine,jaynes57a,jaynes57b,jaynes80,paltridge75,paltridge78,paltridge79,paltridge81,paltridge01,dewar03,vilar} which can be used for local solutions when the system is not far from equilibrium and dissipative processes dominate there, there is no consensus in if these principles can be applied to mesoscopic transport~\cite{bokes03}.

The mesoscopic transport should have been a touchstone of the principles in nonequilibrium statistical mechanics. But it is not the case because of lack of definitions of local intensive thermodynamic variables, which play central roles in statistical mechanics. In evolution approach even if the currents can be easily calculated, up to now there is no way of calculating the temperature in a local region of the mesoscopic device. Because the current and the temperature are two different kinds of physical quantities. The current is calculated as the expectation value of the quantum operator of current with respect to the density matrix of the system. While the temperature is intensive, which in statistical mechanics is introduced by hand as a Lagrangian multiplier before one writes down the density matrix. There is no ``temperature operator'', although in fact temperature is also measurable like all the other physical quantities. Such a discrepancy between temperature and other physical quantities causes troubles in the theory of quantum transport. In fact there is no reason for ignoring the local intensive variables in mesoscopic device, since in principle we could continuously increase the device size without meeting a breakdown point of the physical laws behind, i.e. the Schr\"{o}dinger equation, until the device is in the macroscopic region where the driving forces are the gradients of intensive variables rather than their differences in reservoirs.

In this paper we get rid of this discrepancy by introducing a method of checking if the local equilibrium condition is fulfilled in arbitrary nonequilibrium steady states and of calculating the local temperature and chemical potential. We apply our method in a disordered chain subject to a current, and study the onsite chemical potential by attaching an auxiliary site to it. We study the distribution of onsite chemical potentials at different disorder strength and reservoir voltage bias. In Sec.~II we discuss how to reach the nonequilibrium steady state in an evolution approach by taking thermodynamic limit before taking $t\to \infty$. In Sec.~III, we discuss the condtion of local equilibrium and suggest a method of calculating the local intensive variables. In Sec.~IV, we introduce the model of disordered 1D chain and show the distribution of onsite chemical potentials inside it subject to finite voltage bias. Sec.~V is a short summary.

\section{when will a system evolve into a nonequilibrium steady state}

When we say that a system evolves into a nonequilibrium steady state, we mean that when time goes to infinity the system relaxes towards a stationary state in which some flows are nonzero, but there is no time variation. The NESS should be distinguished with the equilibrium state in which some nonzero flows exist in a ring structure. The latter can exist in an isolated system with finite number of degrees of freedom, such as the mesoscopic ring in a magnetic field with a persistent current. But an isolated system with finite number of degrees of freedom will never evolve into NESS. Since if there exists a flow in an irreversible process, the flow carries some quantity (mass, energy or charge) from one part of the system to the other part, causing a continuous decreasing of the quantity in one part and increasing in the other part due to the conservation law. This contradicts the steady state argument. 

To avoid the paradox between irreversibility and stationarity, we embed the system in an environment so that the composite isolated system is infinite. Then the total quantity in the whole system is infinite, so that any global conservation law is invalidated. At the same time if we see the local subsystem, the density of matter and the flow are both time invariant.

In practice, to approach a nonequilibrium steady state we generally begin with a finite model and then increase the number of degrees of freedom into infinity. This process is called taking thermodynamic limit. One must notice that in most cases taking thermodynamic limit and taking $t\to\infty$ are not exchangeable. To get the nonequilibrium steady state one must take thermodynamic limit before taking $t \to \infty$. Next we give two examples, namely single impurity Anderson model and an infinite chain. 

\subsection{Single impurity Anderson model}

Let us recall single impurity Anderson model which describes a quantum dot coupled to two leads (the left and right leads). This model is exactly solvable without considering electron-electron interaction and has been extensively studied in condensed matter community. It is well known that a stationary current through the dot will be established if the two leads are in different chemical potentials and the coupling between leads and the dot has been switched on in the infinite past. 

The crucial condition for a stationary nonzero current is that the number of levels in leads must be infinite. Otherwise the current perpetually oscillates with an average value of zero. This is easy to see if we suppose there is only one level in each lead. Then the Hamiltonian is written as
\begin{eqnarray}
\hat H=\omega(\hat c^\dag_L \hat c_L+ \hat c^\dag_R \hat c_R+ \hat d^\dag \hat d)+g(\hat c^\dag_L \hat d+ \hat c^\dag_R \hat d+h.c.),
\end{eqnarray}
where $\hat c_L$, $\hat c_R$ and $\hat d$ denote the annihilation operators in left lead, right lead and the dot respectively. The left lead is occupied by an electron at time $t=0$. After switching on the coupling, we find the current to be
\begin{eqnarray}
 I_L(t) = \frac{\sqrt{2}}{2} |g| \sin \left(\sqrt{2} |g|t \right) + \frac{\sqrt{2}}{4} |g| \sin \left(2\sqrt{2} |g|t \right),
\end{eqnarray}
which satisfies the conservation law 
\begin{eqnarray}\label{conservation}
I_L(t)= -\frac{dN_L(t)}{dt}.
\end{eqnarray}
Here $N_L$ is the electron number in left lead. Whatever finite number of levels there are in leads, the time-averaged current must be zero. Otherwise the current will eventually empty one lead and overflow the other. In thermodynamic limit, however, the term in righthand side of Eq.~\ref{conservation} is nonsense since $N_L$ is infinite. So the conservation law does not prohibit a nonzero current any more. 

One should notice that the stationary current in NESS is not carried by the eigenstates of the Hamiltonian. The invariant term of current operator in Anderson model is zero. In other words, the expecation value of $\hat I_L$ is zero with regard to arbitrary eigenstate, because the current operator is $\hat I_L \sim [\hat H,\hat N_L] $. The fact that the eigenstates carry no current distinguishes NESS with equilibrium states carrying nonzero current.

\subsection{An infinite chain}

In single impurity Anderson model we employ the open boundary conditions, while the NESS can also be approached in an infinite chain with periodic boundary conditions. 

The Hamiltonian of a chain of length $2N$ with periodic boundary conditions is written as 
\begin{eqnarray}\label{hoppinghamiltonian}
\hat H= -g \sum_{i=1}^{2N-1} (\hat c^\dag_i \hat c_{i+1}+h.c.) -g (\hat c^\dag_{2N} \hat c_1+h.c.).
\end{eqnarray}
In momentum basis it is diagonal as $\hat H=\sum_{k=0}^{2N-1} \epsilon_k \hat c^\dag_k \hat c_k$ where $\epsilon_k= -2g \cos (2\pi k/2N)$. The operator at site $j$ is expressed in momentum basis as
\begin{eqnarray}
\hat c_j = \sum_{k=0}^{2N-1} \frac{e^{i2\pi jk/2N}}{\sqrt{2N}} \hat c_k.
\end{eqnarray}
Then we immediately have 
\begin{eqnarray}\nonumber
\hat c_j(t)= \sum_{j'=1}^{2N} W_{jj'}\hat c_{j'},
\end{eqnarray}
where $W_{jj'}= \sum_{k=0}^{2N-1} \displaystyle \frac{1}{2N}e^{i(\frac{\pi k(j-j')}{N}-\epsilon_k t)}$ is the propagator. 

Now we suppose that the sites from $j=1$ to $j=N$ are all occupied by electrons and sites from $j=N+1$ to $j=2N$ are empty at initial time $t=0$. Then the hopping interaction in Eq.~\ref{hoppinghamiltonian} is switched on. The local current and electron density at arbitrary time are worked out easily, given the current operator from site $j$ to $(j+1)$ as
\begin{eqnarray}
\hat I_{j\to j+1} = -ig (\hat c^\dag_j \hat c_{j+1}- \hat c^\dag_{j+1} \hat c_j).
\end{eqnarray}
Obviously this definition satisfies the local conservation law $\frac{d n_j}{dt} = I_{j-1\to j} -I_{j\to j+1}$. The current evaluates to
\begin{eqnarray}
 I_{j\to j+1}(t)= 2g \sum_{j'=1}^N \textbf{Im} W^*_{jj'}W_{j+1,j'},
\end{eqnarray}
and the electron density at site $j$ evaluates to
\begin{eqnarray}
 n_j(t)= \sum_{j'=1}^N |W_{j,j'}|^2.
\end{eqnarray}
According to the behavior of scaled parameter $g$ as $N \to \infty$, the system will evolve into either a NESS with nonzero local current or a steady state with zero current everywhere.

\begin{figure}
\includegraphics[width=0.45\textwidth]{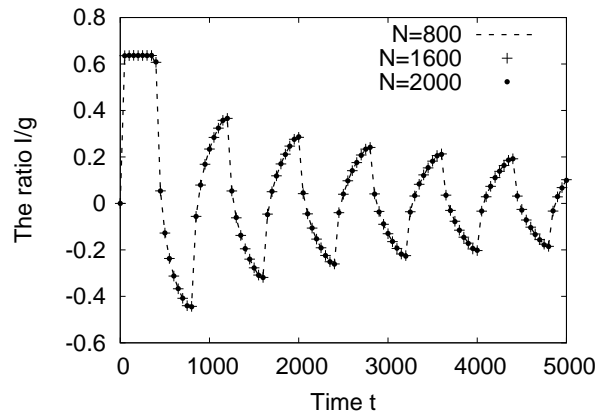}
\caption{The evolution of current from site $N$ to $N+1$ for different chain length when $g/N$ is a constant. Here we set $g=N/800$. The $y$-axis is the ratio of current to $g$ and $x$-axis the time. One should notice that the unit of time is not shown in the figure. But it is in fact $800N/g$. The function $I_{N\to N+1}(t)/g$ has converged when $N$ is as large as $800$.}
\end{figure}
If $g/N$ keeps a constant, the electron density $n_j(t)$ and the ratio $I_{j\to j+1}(t)/g$ oscillate around some values with a period which does not vary with $N$ (see Fig.~1). As $N\to \infty$, $n_j(t)$ and $I_{j\to j+1}(t)/g$ both have well defined thermodynamic limits. For arbitrary $j$, the oscillation amplitudes of $n_j(t)$ and $I_{j\to j+1}(t)/g$ decay in course of time. When time goes to infinity the electron density approaches to $0.5$ and the current to zero everywhere. 

If $g$ keeps a constant, however, as $N$ increasing the period of functions $n_j(t)$ and $I_{j\to j+1}(t)$ increases accordingly. For any finite $N$, the current eventually decays towards zero. But if we take thermodynamic limit firstly, the period of oscillations goes to infinity. The current at some sites will approach to a nonzero value when we take $t\to \infty$ thereafter. 
\begin{figure}
\includegraphics[width=0.45\textwidth]{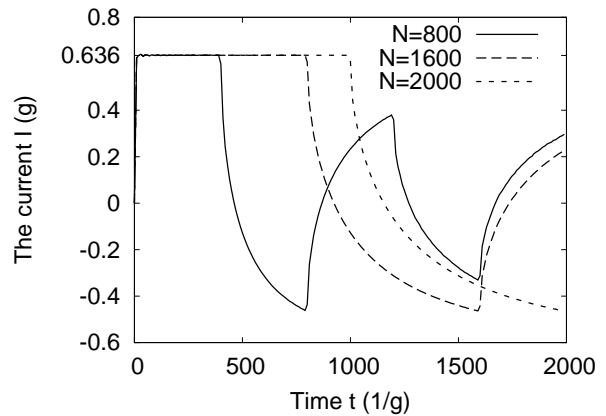}
\caption{The evolution of current for different $N$ when $g$ is a constant. Here we set $g=1$. With increasing $N$ the period of current will go to infinity. If one takes $N\to \infty$ before taking $t\to\infty$, the steady limit of current will be around $0.636$.}
\end{figure}

Let us first study the currents at sites $j=(1+x)N$ as $0<x<1$ is a constant. One should notice $j$ will vary with $N$. In thermodynamic limit these sites are infinitely far away from the electron reservoir (the sites from $1$ to $N$), so that the current $I_{j\to j+1}$ and the electron density $n_j$ both approach to zero. On the other hand at the site $j=(1+x)N$ as $-1<x<0$, which is in the reservoir, the current approaches to zero and the electron density approaches to one. 

The site $N+n$ as $n$ is an arbitrary integer is in the boundary between the reservoir and the vacuum. There the current approaches approximately to $0.636$ (see Fig.~2), a nonzero value. At the same time the electron density approaches to $0.5$. Here $n$ can be arbitrarily large or small since in thermodynamic limit we always have $|n| \ll N$ whatever $n$ is. Then in the system consisting of sites $N+n$, which can be even macroscopic in size, there exists a well defined nonequilibrium steady state. 

As shown above, solving equations of motion and taking thermodynamic limit before taking $t\to\infty$ supply a platform for studying the NESS. In next section we will discuss how to define the thermodynamic intensive variables, such as temperature and chemical potential, in the NESS got by this approach. The definition of intensive variabls is important both in experiments where they can be measured and in the theory of nonequilibrium thermodynamics where they must be given before hand.

\section{how to calculate local temperature and chemical potential in a system in nonequilibrium steady states}

Today quantum mechanics is believed to be the universal theory describing the evolution of both microscopic and macroscopic systems. In quantum mechanics the observables are represented by linear operators acting on the Hilbert space. On the other hand the concepts in themodynamics, such as temperature $T$ and chemical potential $\mu$, have been widely used to describe the macroscopic systems in everyday life. But unfortunately these intensive variables in thermodynamics cannot be directly related to quantum operators. Instead, in statistical mechanics $T$ and $\mu$ are introduced by hand as the Lagrangian multipliers for the energy $\hat H$ and the particle number $\hat N$ respectively. The lack of interpretations for $T$ and $\mu$ at the microscopic level causes problems in nonequilibrium thermodynamics, where there is no generally accepted variational principle and then the temperature and chemical potential (or more strictly the local temperature and local chemical potential) cannot be determined by the density matrix of the system. In other words there is no way of calculating $T$ and $\mu$ in a NESS which is not the result of variational principle but the steady limit of the state evolving according to Schr\"{o}dinger equation. We contribute to solve this problem by providing a method of calculating the local temperature and chemical potential in a system according to its density matrix.

To avoid ambiguity we would like to emphasize that we do not define local temperature and chemical potential in arbitrary NESS. In fact the definitions of $T$ and $\mu$ make sense only in systems where the local thermodynamic equilibrium conditions are fulfilled. In other words, if we divide the system into small cells each cell must look like being in an equilibrium state, even if the whole system undergoes an irreversible process. In systems violating the local equilibrium arguments one would get arbitrary $T$ and $\mu$ by employing different measurement schemes which give unique result in an equilibrium system. As will be shown, an approach examining if the local equilibrium conditions are fulfilled naturally arises from our method.

Let us recall the process of using a mercury thermometer to measure the temperature of an object. We attach the thermometer closely to the target and wait for long enough until the temperatures of the target and the thermometer are the same. If the target is much larger than the thermometer, its temperature will not be changed during the equilibration process. Then we read out the temperature according to the volume of mercury. This procedure suggests a general scheme of measuring local $T$ and $\mu$ of a system either in equilibrium or out of equilibrium by attaching an auxiliary apparatus to the region we want to measure. After waiting for a period longer than the equilibration time, the auxiliary apparatus will equilibrate if the conditions of local equilibrium are satisfied even if the whole system may be still out of equilibrium. Then $T$ and $\mu$ are read out in the auxiliary apparatus according to their orthodox definitions in equilibrium. The auxiliary apparatus (the ``thermometer'') should satisfy the following conditions: it must be very small so that it almost has no effect on the original state of the system; and there is a simple relation between its temperature and some measurable property. Obviously a single site is a suitable candidate of such auxiliary apparatus, since it is the smallest apparatus and its occupation number for fermions can be expressed as 
\begin{eqnarray}\label{occupation}
n_d =\displaystyle \frac{1}{e^{\beta(\epsilon_d -\mu)}+1},
\end{eqnarray}
where $\epsilon_d$, $\beta$ and $\mu$ are the energy level, the inverse of temperature and the chemical potential respectively.

Our method bases on an ambitious assumption that a single site will equilibrate after attached to a large system. This is true only under special conditions, i.e., the coupling between the site and the system is infinitesimal. Consider an impurity site coupled to a Fermi sea with chemical potential $\mu$ and temperature $T$. This is just the single impurity Anderson model without interaction. We switch on the coupling at time $t=0$. Given the energy level $\epsilon_d$ and the level width $\Gamma$ at the impurity site, in steady limit when time goes to infinity we find the occupation number to be
\begin{eqnarray}
 n_d = \frac{\Gamma}{\pi} \int^\infty_{-\infty} d\epsilon \frac{1}{e^{\beta(\epsilon-\mu)}+1} \frac{1}{(\epsilon-\epsilon_d)^2+\Gamma^2}.
\end{eqnarray}
Only in weak coupling limit $\Gamma \to 0$ we have $\frac{\Gamma}{\pi} \frac{1}{(\epsilon-\epsilon_d)^2+\Gamma^2} \to \delta(\epsilon-\epsilon_d) $ and then recover Eq.~\ref{occupation}. We rewrite Eq.~\ref{occupation} as 
\begin{eqnarray}\label{linearrelation}
 \ln \left(\frac{1}{n_d}-1 \right)= \beta \epsilon_d -\beta \mu .
\end{eqnarray}
Then $\beta$ and $\mu$ can be easily determined by plotting $ \ln (1/n_d-1)$ with respect to $\epsilon_d$. 

If the condition of local thermodynamic equilibrium is satisfied, according to the definition, the auxiliary site attached to the system must be in equilibrium and then Eq.~\ref{linearrelation} must be satisfied. Conversely, if the linear relation in Eq.~\ref{linearrelation} is satisfied at the attached site, we would make sure that the condition of local equilibrium is fulfilled.

In summary, we let a system evolve into a nonequilibrium steady state. Then we couple an auxiliary site to a local region of the system. The coupling strength should be infinitesimal. We switch on the coupling and calculate the occupation number $n_d$ at the site in steady limit. We adjust the energy level $\epsilon_d$ at the auxiliary site to obtain a series of pairs $(\epsilon_d,n_d)$. If we cannot fit Eq.~\ref{linearrelation} to the series of data points, we conclude that the local region is not in equilibrium. If we succeed, we then determine the local temperature and chemical potential from the fitting parameters. This is our method of using one auxiliary site to calculate the local intensive thermodynamic parameters in a NESS.

Next we give some comments on the possible applications of it. Up to now solving the equation of motion is the only generally accepted way to reach a NESS in mesoscopic transport. In this approach we usually assume a central conductor connected to at least two infinite reservoirs which are in equilibrium at initial time. The boundary conditions of the central conductor are given by the temperatures and chemical potentials of the reservoirs, which are assumed not to vary with time. The steady current through the central conductor can be calculated. The curve of current vs. voltage bias is plotted and compared with experimental results. However, sometimes we are more interested in the distribution of local chemical potentials inside the central scattering region, which cannot be resolved from boundary conditions. Our method provides a way to calculate the local $T$ and $\mu$ in the central conductor. The results can be compared with experiments. The local $T$ and $\mu$ are more important than reservoir temperature and chemical potential in the theory of nonequilibrium thermodynamics, where the gradients of intensive variables are called thermodynamic forces and are the reasons of flows.

\section{The distribution of chemical potentials inside a disordered chain}

A chain connected to two reservoirs has been widely used to model the quantum point contact and the quantum wire fabricated in semiconductor heterostructures. If the two reservoirs are in different chemical potentials, a current will be driven through the wire. This is the Landauer-B\"{u}ttiker picture for mesoscopic transport. In ballistic transport regime when the wire is clean, there is only contact resistance in the device. In other words, there are sharp voltage drops at the interface between the reservoirs and the wire, while the voltage is equal everywhere inside the wire. 

What is more interesting is a dirty wire, where we would expect the intrinsic resistance and a gradual drop of voltage through the wire. As in a macroscopic circuit, in a dirty wire the current can be treated as the result of the gradient of voltage inside the wire instead of the voltage bias of the reservoirs. The reservoirs are no more than the simulation of the voltage source in a circuit, while they are necessary for preparing the NESS. 

A disordered chain is obtained by proposing random onsite potentials $\epsilon_j$, where $j$ denotes the chain site. For simplicity we suppose that there is no correlation between $\epsilon_j$ at different sites. The average of $\epsilon_j$ is denoted as $\bar \epsilon$. We suppose that $\epsilon_j$ has a uniform distribution at the interval $[\bar\epsilon-\frac{\Delta}{2},\bar\epsilon+\frac{\Delta}{2}]$, where $\Delta$ denotes the disorder strength. Then the total Hamiltonian including the two reservoirs is written as
\begin{eqnarray}\nonumber\label{interactinghamiltonian}
\hat H &=& \sum_{k\alpha\sigma} \epsilon_k \hat c^\dag_{k\alpha\sigma} \hat c_{k\alpha\sigma} + g_l \sum_{k\sigma} ( \hat c^\dag_{kL\sigma} \hat c_{1\sigma} + \hat c^\dag_{kR\sigma} \hat c_{n\sigma} +h.c.) \\ && \nonumber + \sum_{\sigma,j=1}^n \epsilon_j \hat c^\dag_{j\sigma} \hat c_{j\sigma} + g \sum_{j=1}^{n-1} ( \hat c^\dag_{j\sigma} \hat c_{j+1,\sigma} + h.c.) \\ && + U\sum_{j=1}^n \hat c^\dag_{j\uparrow} \hat c_{j\uparrow} \hat c^\dag_{\downarrow} \hat c_{j\downarrow} ,
\end{eqnarray}
where $\hat c_{k\alpha\sigma}$ is the annihilation operator of the electron in the lead, $\sigma=\uparrow,\downarrow$ denotes the spin and $\alpha=L,R$ denotes the left and right lead. The leads are in thermal equilibrium at temperature $T_L$ and $T_R$, and in chemical potential $\mu_L$ and $\mu_R$ respectively. The operator $\hat c_{j\sigma}$ with $j=1,2\cdots n$ denotes the annihilation operator at site $j$ in the chain. The coupling strength between nearby sites in the chain is set to be $g$, and that between the leads and the chain is set to be $g_l$. 

For general parameters, the above Hamiltonian is non-integrable due to the existence of Coulomb interaction. Several complicated numerical methods and approximation schemes have been developed for solving this model. For simplicity, in this paper we will first solve the model without considering interaction by setting $U=0$. Then we consider the effect of the interaction in self-consistent mean field approximation. Without interaction the electrons in different spin channels will transport independently and the spin index can be neglected. The simplified Hamiltonian becomes
\begin{eqnarray}\nonumber\label{chainhamiltonian}
\hat H &=& \sum_{k,\alpha=L,R} \epsilon_k \hat c^\dag_{k\alpha} \hat c_{k\alpha} + g_l \sum_k ( \hat c^\dag_{kL} \hat c_{1} + \hat c^\dag_{kR} \hat c_n +h.c.) \\ && + \sum_{j=1}^n \epsilon_j \hat c^\dag_j \hat c_j + g \sum_{j=1}^{n-1} ( \hat c^\dag_j \hat c_{j+1} + h.c.) .
\end{eqnarray}

In the left and right electron reservoirs we employ the momentum basis and assume a constant density of states denoted by $\rho$. At the same time the reservoirs can also be simulated by the semi-infinite chains. In this sense the Hamiltonian of Eq.~\ref{chainhamiltonian} describes an infinite chain with emphasized central sites, similar to the Hamiltonian of Eq.~\ref{hoppinghamiltonian}. If we define $\Gamma =\rho \pi g_l^2$ as the energy unit (like what we do in single impurity Anderson model) and set $g=\Gamma$, the coupling strength between central sites has the same amplitude as that between nearby sites in reservoirs. This corresponds to a comparatively strong coupling between central sites. In following text we always set $g=\Gamma$. One could also take different values of $g$. But as $g\ll \Gamma$ the transmission spectrum has very sharp peaks, and one should be very careful when performing numerical integration routines. 

\subsection{The electron density and the current at site $m$}
The steady current and electron density at arbitrary site $m$ in the disordered chain are got by the Keldysh technique, in which an adiabatic evolution is proposed with $g$ and $g_l$ switched on little by little. In this model adiabatically switching on the coupling makes no difference with a quench of coupling for the physical quantities in steady limit. In details we define the retarded Green functions as
\begin{eqnarray}\label{retardeddef}
 G^r_{i,j}(t,t')=-i \theta(t-t') \langle \{ \hat c_i(t), \hat c_j^\dag(t')\} \rangle, 
\end{eqnarray}
and the lesser Green functions as
\begin{eqnarray}\label{deflesser}
 G^<_{i,j}(t,t')=i \langle \hat c_j^\dag(t') \hat c_i(t)\rangle, 
\end{eqnarray}
where $i,j=1,2,\cdots n$ denote the sites in the chain. The electron density and current at arbitrary site in the chain can be related to the simultaneous lesser Green functions.

The freqency representation of Green functions is defined as
\begin{eqnarray}
G^{r,<}_{i,j}(\omega)=\int d(t-t') e^{i\omega(t-t')} G^{r,<}_{i,j} (t,t'). 
\end{eqnarray}
In frequency representations the retarded Green function satisfies the Dyson equation, i.e.
\begin{eqnarray}\label{dyson}
 G^r(\omega)= G^{0r}(\omega)+ G^{0r}(\omega) \Sigma^r(\omega) G^{r}(\omega),
\end{eqnarray}
where $G^r(\omega)$ is a $n\times n$ matrix whose elements are $G^{r}_{i,j}(\omega)$. $G^{0r}(\omega)$ is the free Green function matrix when $g=g_l=0$, and its elements are 
\begin{eqnarray}\label{g0rdef}
{G}^{0r}_{i,j}(\omega) = \delta_{i,j} \frac{1}{\omega- \epsilon_j+i\eta},
\end{eqnarray}
where $\eta>0$ is infinitesimal. $\Sigma^r(\omega)$ is the self-energy matrix, which comes from two sources. The first is the hopping energy between nearby sites in the chain, i.e.
\begin{eqnarray}\label{tridiagonalelement}
\Sigma^r_{j,j+1} = \Sigma^r_{j+1,j} = g. 
\end{eqnarray}
The second is the broadening of levels at the edge sites due to the coupling to leads, i.e.
\begin{eqnarray}\nonumber\label{edgeelement}
 \Sigma^r_{1,1}(\omega) &=&  \Sigma^r_{n,n}(\omega) \\ &=& \nonumber g^2_l \sum_k \frac{1}{\omega- \epsilon_k +i\eta} \\ &=& -i\Gamma, 
\end{eqnarray}
where $\Gamma = \rho \pi g_l^2$. The other elements of the self-energy matrix are all zero. By sloving Eq.~\ref{dyson} we express the retarded Green function as
\begin{eqnarray}\label{solvedyson}
 G^r(\omega)= \left(G^{0r}(\omega)^{-1}-\Sigma^r(\omega) \right)^{-1}.
\end{eqnarray}
Here we need to calculate the inverse of a $n\times n$ matrix, which can be finished by computer.

The lesser Green function in frequency representation is related to the retarded one in Keldysh formalism, i.e.
\begin{eqnarray}\label{lesserpreresult}
 G^<_{i,j}= g^2_l\left( G^r_{i,1}\sum_k G^{0<}_{kL} G^a_{1,j} + G^r_{i,n} \sum_k G^{0<}_{kR} G^a_{n,j} \right),
\end{eqnarray}
where $G^{0<}_{k\alpha}(\omega)= 2\pi i \displaystyle \frac{1}{e^{\beta_\alpha(\omega-\mu_\alpha)}+1} \delta(\omega-\epsilon_k)$ with $\alpha=L,R$ is the lesser Green function in left and right leads respectively. Here $\beta_\alpha=1/(k_B T_\alpha)$ and $\mu_\alpha$ are the inverse of temperature and the chemical potential in lead $\alpha$ respectively, and $G^a_{i,j}$ denotes the advanced Green function, which is the complex conjugation of the corresponding retarded Green function, i.e. $G^a_{i,j}=(G^r_{j,i})^*$. Eq.~\ref{lesserpreresult} evaluates
\begin{eqnarray}\label{lesserresult}
 G^<_{i,j}= 2i\Gamma \left(G^r_{i,1} G^a_{1,j}f_L+G^r_{i,n} G^a_{n,j}f_R\right),
\end{eqnarray}
where $f_\alpha(\omega)=\displaystyle\frac{1}{e^{\beta_\alpha(\omega-\mu_\alpha)}+1}$ is the Fermi function in lead $\alpha$. 

At last we perform inverse Fourier transformation and get the lesser Green function in time representation, i.e.,
\begin{eqnarray}\label{inversefourier}
G^<_{i,j}(t,t')= \int_{-\infty}^\infty d\omega \frac{e^{-i\omega(t-t')}}{2\pi} G^<_{i,j}(\omega).
\end{eqnarray}
The electron density at site $m$ is defined as $n_m=\langle \hat c^\dag_m \hat c_m \rangle$, which can be expressed as $n_m= -iG^<_{m,m}(0,0)$ according to the definition of the lesser Green function. Substituting Eq.~\ref{inversefourier} in, we have
\begin{eqnarray}\label{electrondensity}
 n_m = \frac{1}{2\pi i} \int d \omega G^<_{m,m}(\omega).
\end{eqnarray}
The current from site $m$ to $(m+1)$ is expressed as $ I_{m\to m+1}= \displaystyle \frac{g}{\hbar} \langle i (\hat c^\dag_m \hat c_{m+1} -h.c.) \rangle$, which satisfies the electron number conservation law $\displaystyle \frac{d n_m}{dt}= I_{m-1\to m}- I_{m\to m+1}$. Similarly it is related to the lesser Green function as
\begin{eqnarray}\label{currentatsite}
 I_{m\to m+1} = \frac{-2g}{h} \textbf{Re} \int d \omega G^<_{m,m+1}(\omega).
\end{eqnarray}
Eq.~\ref{solvedyson},~\ref{lesserresult},~\ref{electrondensity} and~\ref{currentatsite} together give the complete procedure for calculating the electron density and current at arbitrary site.

In numerical approach, we first use a random number generator to generate a group of onsite potentials $\epsilon_j$. One should notice that the distribution of $\epsilon_j$ is uniform at the interval $[\bar\epsilon-\frac{\Delta}{2},\bar\epsilon+\frac{\Delta}{2}]$. Then the electron density and current are calculated subject to this group of $\epsilon_j$. Obviously repeating the calculation will not produce the same result, since $\epsilon_j$ is randomly generated. We are more interested in the average value of current, which is got by repeating the calculation for many times. In each time a new group of $\epsilon_j$ is generated. We repeat this procedure until the averaged current is convergent. 

\subsection{The onsite temperature and chemical potential}

We calculate the temperature and chemical potential at arbitrary site in the chain. This is done by attaching an auxiliary site to the site we want to measure. For example, if we want to measure the temperature $T_m$ and the chemical potential $\mu_m$ at site $m$, we will modify the Hamiltonian of Eq.~\ref{chainhamiltonian} by including the extra terms
\begin{eqnarray}
\hat H_{mea}= g'(\hat d^\dag \hat c_m+h.c.) + \epsilon_d \hat d^\dag \hat d,
\end{eqnarray}
where $\hat c_m$ and $\hat d$ denote the annihilation operators at site $m$ and the auxiliary site respectively. Here $\epsilon_d$ denotes the energy level at the auxiliary site and is adjustable, and $g'$ denotes the auxiliary coupling strength which should be infinitesimal. The whole system consisting of the chain coupled to two leads and the auxiliary site is described by the Hamiltonian $\hat H+\hat H_{mea}$, where $\hat H$ is defined in Eq.~\ref{chainhamiltonian}.

Again the Keldysh technique is employed to calculate the occupation number at the auxiliary site $n_d=\langle \hat d^\dag \hat d \rangle$ corresponding to different $\epsilon_d$. This should be done by adiabatically switching on $g'$ after the chain described by $\hat H$ has been in nonequilibrium steady state. Due to lack of initial correlation, this is equivalent to switching on $g_l$, $g$ and $g'$ simultaneously when we are only interested in the steady limit of the local quantity $n_d$.

Now the chain and the auxiliary site together are treated as the central conductor, which totally contains $(n+1)$ sites. We define the retarded Green function concerning the auxiliary site as
\begin{eqnarray}\nonumber
 G^r_{i,n+1}(t,t')&=& -i \theta(t-t') \langle \{ \hat c_i(t), \hat d^\dag(t')\} \rangle, \\ \nonumber
 G^r_{n+1,j}(t,t')&=& -i \theta(t-t') \langle \{ \hat d(t), \hat c_j^\dag(t')\} \rangle, \\
 G^r_{n+1,n+1}(t,t')&=& -i \theta(t-t') \langle \{ \hat d(t), \hat d^\dag(t')\} \rangle,
\end{eqnarray}
where $i,j=1,2,\cdots n$. Together with the Green functions defined in Eq.~\ref{retardeddef} we get a $(n+1)\times (n+1)$ matrix. Similar to Eq.~\ref{solvedyson} it can be expressed as
\begin{eqnarray}
G^r(\omega)= \left(G^{0r}(\omega)^{-1}- \tilde \Sigma^r(\omega) \right)^{-1},
\end{eqnarray}
where $G^{0r}(\omega)$ is the $(n+1)\times (n+1)$ diagonal matrix. The elements of $G^{0r}$ are defined in Eq.~\ref{g0rdef} except for the one concerning the auxiliary site $G^{0r}_{n+1,n+1}= \displaystyle \frac{1}{\omega- \epsilon_d+i\eta}$. Now the self-energy matrix comes from three sources: the hopping energy between nearby sites in the chain; the coupling energy between the edge sites and the leads; and the coupling energy between $m$ site and the auxiliary site. The corresponding elements are
\begin{eqnarray}
\tilde \Sigma^r_{j,j+1} = \tilde\Sigma^r_{j+1,j} = g
\end{eqnarray}
with $j=1,2,\cdots,n-1$,
\begin{eqnarray}
\tilde \Sigma^r_{1,1} = \tilde\Sigma^r_{n,n} = -i\Gamma,
\end{eqnarray}
and
\begin{eqnarray}
\tilde \Sigma^r_{m,n+1} = \tilde\Sigma^r_{n+1,m} = g'.
\end{eqnarray}

The expression of the lesser Green function in Eq.~\ref{lesserresult} keeps invariant because the auxiliary site is not directly coupled to the leads. Especially we have
\begin{eqnarray}
 G^<_{n+1,n+1}= 2i\Gamma \left(|G^r_{n+1,1}|^2f_L+|G^r_{n+1,n}|^2 f_R\right).
\end{eqnarray}
The occupation number at the auxiliary site is expressed as
\begin{eqnarray}
 n_d = \frac{1}{2\pi i} \int d \omega G^<_{n+1,n+1}(\omega).
\end{eqnarray}

We have known that $g'\to 0$ is the necessary condition of local equilibrium at the auxiliary site. In numerical approach, the weak coupling limit is realized by self-adaptation algorithm. We begin with a finite $g'$, and reduce it by half in each loop until the final result $n_d$ converges to a desired precision.

At beginning of the numerical approach, we generate a group of onsite potentials $\epsilon_j$. Then we in turn calculate the current through the chain and the occupation number at the auxiliary site with respect to different $\epsilon_d$ and $m$. When calculating the current, we turn off the coupling to the auxiliary site. One should notice that the destiny of an auxiliary site is to measure the temperature and the chemical potential at sites in the chain. One does not measure different sites simultaneously but rather one after another. We finally get a series of functions $n_d(\epsilon_d)$ as $m$ varying from $1$ to $n$. We then calculate the average of temperature and chemical potential at site $m$ with respect to different configurations of disorder.

\subsection{Transition from delocalization to localization as the disorder strength $\Delta$ increasing}

\begin{figure}
\includegraphics[width=0.45\textwidth]{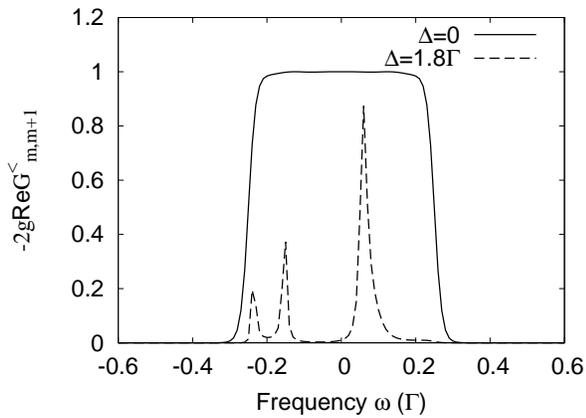}
\caption{The transmission spectrum $-2g\textbf{Re} G^<_{m,m+1}(\omega)$ at different disorder strength. In this figure the chain length is set to be $50$. We set $V=\mu_L-\mu_R=0.5\Gamma$ and $T_L=T_R=0.01\Gamma$.}
\end{figure}
\begin{figure}
\includegraphics[width=0.45\textwidth]{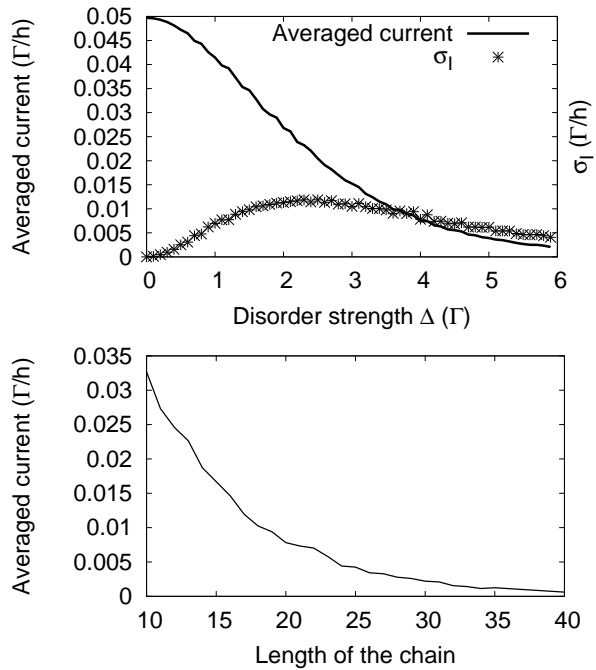}
\caption{The top figure shows the averaged current $\bar I$ and the standard deviation of current $\sigma_I$ as a function of disorder strength $\Delta$. The voltage bias between two leads is $V=\mu_L-\mu_R=0.05\Gamma$, and the temperature in leads is $T_L=T_R=0.1\Gamma$. The length of the chain is set to be $10$. We observe the localization-delocalization transition in the regime of intermediate disorder strength, where the deviation of current reaches its maximum. A large value of $\sigma_I$ indicates a strong fluctuation of current, which often happens near the critical point of a phase transition. The bottom figure shows the averaged current as a function of the chain length. The voltage bias is $V=0.2\Gamma$. }
\end{figure}
What we are studying is the transport through a mesoscopic chain. At particle-hole symmetry we set the average of the Fermi energies in left and right leads to be equal to the average of the onsite potentials in the chain, i.e. $(\mu_L+\mu_R)/2=\bar\epsilon=0$. The linear conductance of the chain in the absence of disorder ($\Delta=0$) reaches the unitary limit $\displaystyle \frac{e^2}{h}$. In the absence of disorder the transport through the chain is well understood under the framework of ballistic transport.

In the presence of disorder, however, we observe the gradual transition from delocalization to localization as the disorder strength increasing. This transition should be attributed to the Anderson localization. While the transition in our model is a ``smooth" one, unlike the phase transition in thermodynamic limit, since the chain consists of only a few of sites. Strictly speaking, there is no disorder-induced phase transition in an infinite one-dimensional system, where an infinitesimal concentration of impurities will cause the exponential decay in the extension of the wave function. In a chain with a few of sites, we see the gradual transition from a continuous band to sharp peaks and wide forbidden regimes between peaks in the transmission spectrum as the disorder strength $\Delta$ increasing (see Fig.~3). This is understood as the transition from ballistic transport in the clean limit to the transport shuttled by the localized levels weakly coupled to the reservoirs. 

The feature of Anderson localization is clearly observed in the figure of averaged current (see the bottom figure of Fig.~4), in which the current shows an exponential decay as the length of the chain increasing. When we calculate the current, the onsite potentials are randomly generated according to $\bar \epsilon$ and $\Delta$. For each generation of onsite potentials we obtain a different current. The averaged current is expressed as
\begin{eqnarray}\label{averagecurrent}
 \bar I = \frac{\sum_{j=1}^M I_j }{M},
\end{eqnarray}
and the standard deviation of current is expressed as
\begin{eqnarray}
 \sigma_I = \sqrt{\frac{\sum_{j=1}^M (I_j-\bar I)^2 }{M}},
\end{eqnarray}
where $I_j$ is the result of current corresponding to the $j$-th generation of onsite potentials and $M$ is the total number of generations. 

The top figure of Fig.~4 shows the average and the standard deviation of current varying with disorder strength. In clean limit as $\Delta=0$, the current approximately approaches to $\frac{e^2}{h} V $ and the standard deviation is exactly zero. As the disorder strength $\Delta$ increasing, the averaged current drops monotonously, while the standard deviation first increases until its maximum and then drops towards zero. At the intermediate disorder strength, $\sigma_I$ reaches its maximum. The strong fluctuation of current in this regime indicates the transition from delocalization to localization. 

\subsection{Local equilibrium condition and the onsite temperature and chemical potential}

\begin{figure}
\includegraphics[width=0.45\textwidth]{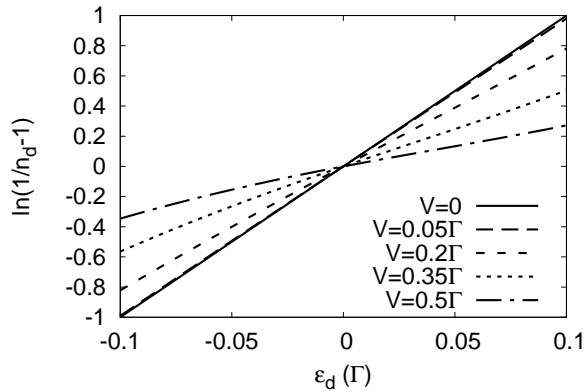}
\caption{An auxiliary site is attached to the third site in a chain of length $10$ coupled to two leads. Whether the condition of local equilibrium is fulfilled at the third site is determined by the relation between the occupation number $n_d$ and the energy level $\epsilon_d$ at the auxiliary site. This figure shows $\ln (1/n_d-1)$ as a function of $\epsilon_d$ at different voltage bias $V=0,0.05\Gamma,0.2\Gamma,0.35\Gamma$ and $0.5\Gamma$. Here we set $\bar \epsilon = 0$ and the disorder strength $\Delta=0$. The temperature in left and right leads is $T_L=T_R=0.1\Gamma$. One should notice that the curves corresponding to $V=0$ and $V=0.05\Gamma$ have strong overlap with each other and are difficult to be distingished. Obviously when the whole system is in thermodynamic equilibrium as $V=0$, the site we are studying is exactly in local equilibrium, as shown in the straight line titled $V=0$. The slope of this line is easily found to be $10/\Gamma$, which is just the inverse of local temperature. The line intercepts the axes at original point, indicating the local chemical potential is zero. If a voltage bias as small as $V=0.05\Gamma$ is applied to the system, the function $\ln (1/n_d-1)$ keeps approximately a linear function and then the condition of local equilibrium is still fulfilled. But as voltage bias increasing, the function $\ln (1/n_d-1)$ obviously deviates from a linear function, as shown in the curve titled $V=0.5\Gamma$.}
\end{figure}
\begin{figure}
\includegraphics[width=0.45\textwidth]{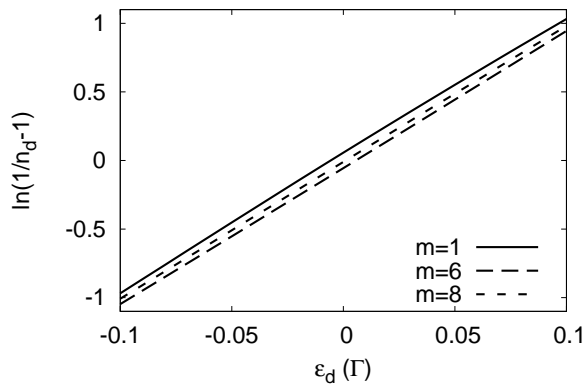}
\caption{The value of $\ln (1/n_d-1)$ is plotted with respect to $\epsilon_d$ at different sites in a disordered chain of length $10$. The left and right lead temperature is $T_L=T_R=0.1\Gamma$. The voltage bias is set to be $V=0.05\Gamma$. The disorder strength is $\Delta=\Gamma$.}
\end{figure}
\begin{figure}
\includegraphics[width=0.45\textwidth]{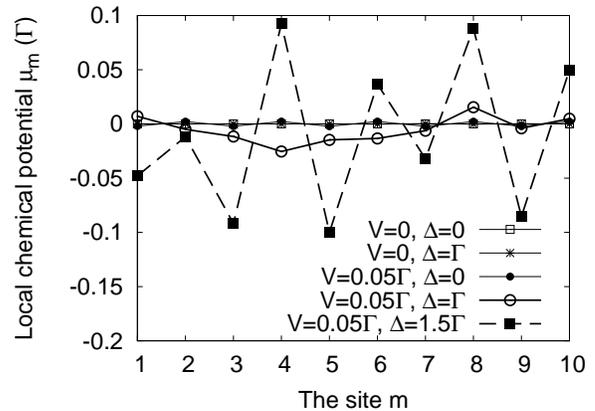}
\caption{The distribution of onsite chemical potentials inside a chain of length $10$ at different voltage bias $V$ and disorder strength $\Delta$. Strong fluctuation of onsite chemical potentials is observed at finite voltage bias in the presence of disorder. }
\end{figure}
\begin{figure}
\includegraphics[width=0.45\textwidth]{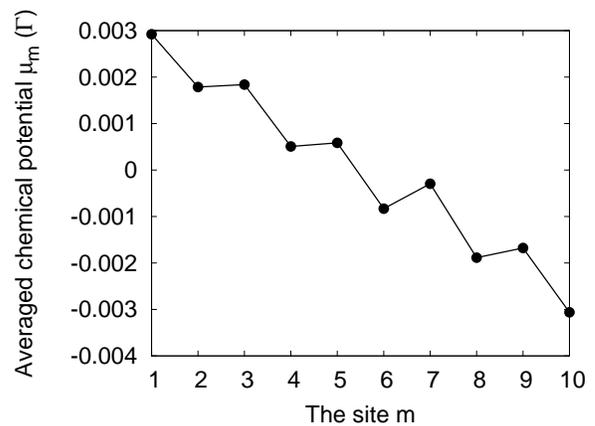}
\caption{The distribution of averaged onsite chemical potentials. Here we set the voltage bias to be $V=0.05\Gamma$ and the disorder strength to be $\Delta=\Gamma$. By taking average with respect to configurations of disorder we observe the regular distribution replacing strong fluctuations.}
\end{figure}

Local thermodynamic equilibrium is an important concept in the study of nonequilibrium thermodynamics. Generally the local equilibrium condition is only fulfilled in the systems not far from equilibrium. In the case of a chain connected to two leads, it requires that the temperature difference and the voltage bias of the two leads are small. We have explained how to use an auxiliary site to check if the local equilibrium condition at arbitrary site in the chain is fulfilled by calculating the occupation number at the auxiliary site as a function of the energy level. In Fig.~5 we plot $\ln(1/n_d-1)$ as a function of $\epsilon_d$ under different voltage bias, which is defined as $V=\mu_L-\mu_R$. According to Eq.~\ref{linearrelation}, the local equilibrium condition is fulfilled if this function is a straight line. As shown in Fig.~5, when $V$ is exactly zero or very small the function of $\ln(1/n_d-1)$ is in fact linear. While as $V$ increasing to as large as $0.5\Gamma$ the function obviously deviates from a linear function. Then the function $n_d(\epsilon_d)$ cannot be regarded as a Fermi function at large voltage bias, and the concepts of temperature and chemical potential are nonsense.

An interesting question arises as to whether the local equilibrium condition is less stringent if the auxiliary site is simultaneously coupled to a large number of sites in the chain (and not only to a given site). We study a model in which the auxiliary site is coupled to up to $100$ sites in a chain of length $200$. Under finite voltage bias, the deviation of the function $\ln(\displaystyle\frac{1}{n_d(\epsilon_d)}-1)$ from a straight line is found to be indepedent to the number of sites measured simultaneously, in opposite to the results observed in an equilibrium system~\cite{hartmann04}. 

At small voltage bias, the function $\ln(1/n_d-1)$ is linear at arbitrary site inside the chain. The slope of the function is explained as the inverse of local temperature, and its value at $\epsilon_d=0$ as $(-\beta \mu)$ where $\mu$ is the local chemical potential. Then we can extract the local temperature $T_m=1/ \beta_m$ and the local chemical potential $\mu_m$ at each site $m$. One may have doubt on the validation of this method. Since what we are interested in is the distribution of onsite temperatures and chemical potentials in nonequilibrium, when they are different from the reservoir ones. However at finite voltage bias, the condition of local equilibrium is only fulfilled approximately. Strictly speaking, the function $\ln(1/n_d-1)$ deviates from a linear function a little bit even at small voltage bias. Then how do we know that the discrepancy between onsite chemical potentials and reservoir ones if there is any is not the result of the deviation of $\ln(1/n_d-1)$ from a linear function? This suspicion is canceled by plotting the function $\ln(1/n_d-1)$ at different sites simultaneously (see Fig.~6). In Fig.~6, we see that subject to a finite voltage bias and disorder strength the functions of $\ln(1/n_d-1)$ at different sites are parallel to each other. They have obviously different intercepts with the axes and at the same time the same slope which is just the inverse of the reservoir temperature. This is a strong evidence that the deviation of $\ln(1/n_d-1)$ from a linear function is very small compared to the difference between onsite potentials. Then our method resolves the onsite chemical potentials to a good extent.

We study the distribution of onsite chemical potentials at different voltage bias and disorder strength. Even if the result depends on the configuration of onsite potentials and is not repeatable when $\Delta \neq 0$, it shows some features which can be attributed to $V$ and $\Delta$ (see Fig.~7). In equilibrium as $V=0$, the onsite chemical potential is zero throughout the chain whether there is disorder or not, coinciding with the fact that the chemical potential in an equilibrium system should be equal everywhere. When there is no disorder, the onsite chemical potential keeps zero everywhere even at finite voltage bias. Because without disorder the resistance of the device comes only from contact resistance, and there is no drop of voltage inside the chain. In the presence of both voltage bias and disorder, we see strong fluctuation of onsite chemical potentials. As disorder strength increasing, the fluctuation becomes stronger. This is contrary to the familiar phenomena in macroscopic circuits where we expect a monotonic drop of voltage through the resistance wire. This can be explained by the mesoscopic nature of the chain, where the movement of electrons must be described by quantum mechanics. In localization regime with intermediate $\Delta$, the position of the electron in the chain is constrained to a small region, then its kinetic energy has strong fluctuation due to uncertainty principle, leading to the strong fluctuation in the onsite chemical potentials.

The abnormal fluctuation of onsite chemical potentials is then canceled by taking the average with respect to configurations of disorder. Similar to the averaged current in Eq.~\ref{averagecurrent}, we define the averaged chemical potential as
\begin{eqnarray}
 \bar \mu_m = \frac{\sum_{j=1}^M \mu_m^j}{M},
\end{eqnarray}
where $\mu_m^j$ denotes the chemical potential at site $m$ in the $j$-th generation. In a real mesoscopic wire, the configuration of disorder is generally not controllable and unknown to the simulators. By taking average with respect to the disorder configurations, one could obtain some universal feature about the distribution of onsite chemical potentials. This is a well-established method in studying disordered systems. As shown in Fig.~8, the averaged chemical potential drops monotonously through the chain, recovering the feature in macroscopic resistance wires. 

\subsection{The effect of interaction}

We have studied the distribution of onsite chemical potentials in a non-interacting disordered chain. A natural question arises as to what is the influence of electron-electron interaction to our results. Next we consider the complete Hamiltonian in Eq.~\ref{interactinghamiltonian} consisting of the onsite Coulomb interaction. We combine it with the measuring term
\begin{eqnarray}
 \hat H_{mea} =g'\sum_\sigma (\hat d^\dag_\sigma \hat c_{m\sigma} +h.c.) + \epsilon_d \sum_\sigma \hat d^\dag_\sigma \hat d_\sigma.
\end{eqnarray}
The two spin channels are considered at the auxiliary site, while the onsite interaction is not. Because if we consider the Coulomb energy, there would be four states with different energies at the auxiliary site. This increases the difficulty in deciding the local temperature and chemical potential. 

Again the Keldysh techniques are employed to calculate the occupation number at the auxiliary site as a function of $\epsilon_d$ in NESS. The Green function can be expressed in a perturbative expansion of $U$. The Green functions of zeroth order in $U$ have been given in above calculations, i.e., in Eqs.~\ref{solvedyson} and \ref{lesserresult}. We take the self-consistent mean field approximation by summing up all the diagrams in the perturbative expansion in which Hartree type self-energies are inserted. This is finished by replacing the onsite potentials $\epsilon_j$ by $\epsilon_j + U n_j$ in the expression of $G^{0r}(\omega)$ in Eq.~\ref{g0rdef}, hence in expressions of $G^r(\omega)$ and $G^<(\omega)$ in Eqs.~\ref{solvedyson} and \ref{lesserresult}. Here $n_j$ is the occupation number at site $j$ and satisfies the self-consistent equation 
\begin{eqnarray}
n_j =  \displaystyle \frac{1}{2\pi i} \int d \omega G^<_{j,j}(\omega).
\end{eqnarray}
Above self-consistent equations can be solved in the iterative method as the interaction strength $U$ is small.
\begin{figure}
\includegraphics[width=0.45\textwidth]{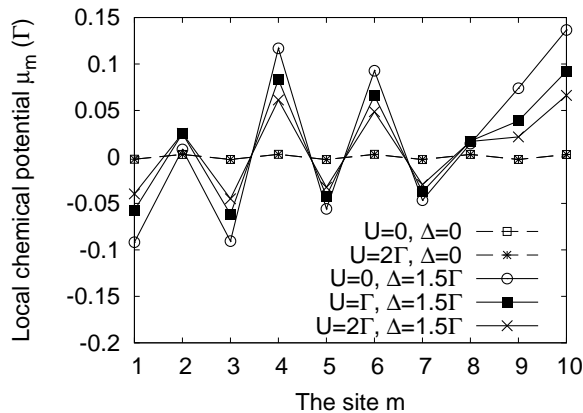}
\caption{The distribution of onsite chemical potentials at different interaction strength $U$ and disorder strength $\Delta$. The chain is set to be at particle-hole symmetry with $\bar\epsilon=-U/2$. The voltage bias is $V=0.05\Gamma$, and the left and right lead temperature are both $0.1\Gamma$. Without disorder the interaction has no effect on the onsite chemical potentials (see the two curves titled $\Delta=0$). In the presence of disorder, the electron-electron interaction will suppress the fluctuation of onsite chemical potentials caused by localization of electrons.}
\end{figure}

We plot the distribution of onsite chemical potentials at the particle-hole symmetric point where $\bar\epsilon=-U/2$ (see Fig.~9). We find that the interaction will suppress the fluctuation of onsite chemical potentials in the presence of disorder. This is similar to the phenomenon that in a closed system the interaction will cause equilibration of the system from arbitrary initial states.

\section{conclusions} 

In summary we study a chain coupled to two semi-infinite leads. Such a chain will evolve into a nonequilibrium steady state after a quench of coupling between leads and the chain if the two leads are in different chemical potentials. We then demonstrate the central idea in this paper of attaching an auxiliary site to the chain to measure the local temperature and chemical potential at each site inside the chain. We find that our method will work if and only if the voltage bias between two leads are small when the conditions of local thermodynamic equilibrium are fulfilled. We then consider a disordered chain by introducing random onsite potentials. The localization-delocalization transition is observed as the disorder strength increasing. More important, we observe that the local onsite chemical potential inside the chain is everywhere zero if either the voltage bias or the disorder is absent. In the presence of both voltage bias and disorder we observe a large fluctuation in onsite chemical potentials, instead of the monotonic drop of voltage usually observed in a macroscopic resistance wire. The fluctuation can be suppressed by the electron-electron interaction. This is attributed to the mesoscopic nature of the chain. By taking average with respect to different configurations of disorder, we recover the monotonic drop of chemical potential. We believe our method can be generalized in arbitrary nonequilibrium steady states to check whether the conditions of local equilibrium are fulfilled and if they are fulfilled to calculate the local intensive thermodynamic parameters.

\end{document}